\documentclass[preprint,showpacs,preprintnumbers,amsmath,amssymb,showkeys]{revtex4}

\usepackage{graphicx}
\usepackage{dcolumn}
\usepackage{bm}

\begin{document}

\title{STATISTICAL ANALYSIS OF MAGNETIC FIELD REVERSALS IN LABORATORY DYNAMO AND IN PALEOMAGNETIC MEASUREMENTS.}

\author{LUCA SORRISO-VALVO$^{1}$, VINCENZO CARBONE$^{1,2}$, MICHAEL BOURGOIN$^{3}$, PHILIPPE ODIER$^{3}$, NICOLAS PLIHON$^{3}$, ROMAIN VOLK$^{3}$}
\affiliation{$^{1}$Liquid Crystal Laboratory, Istituto Nazionale per la Fisica della Materia, Consiglio Nazionale delle Ricerche, ponte P. Bucci, cubo 31C, I-87036 Rende (CS), Italy\\
sorriso@fis.unical.it}

\affiliation{$^{2}$Dipartimento di Fisica, Universit\`a della Calabria, and LICRYL-INFM/CNR, ponte P. Bucci, cubo 31C, I-87036 Rende (CS), Italy\\
carbone@fis.unical.it}

\affiliation{$^{3}$LEGI, Universit\'e Joseph Fourier, 1025 rue de la Piscine, 38000 Grenoble, France\\
mickael.bourgoin@hmg.inpg.fr}

\affiliation{$^{4}$Laboratoire de Physique de l'Ecole Normale Sup\'erieure de Lyon, CNRS UMR5672, 46 All\'ee d’Italie, 69007 Lyon, France}

\begin{abstract}
Statistical properties of the temporal distribution of polarity reversals of the geomagnetic field are commonly assumed to be a realization of a renewal Poisson process with a variable rate. However, it has been recently shown that the polarity reversals strongly depart from a local Poisson statistics, because of temporal clustering. Such clustering arises from the presence of long-range correlations in the underlying dynamo process. Recently achieved laboratory dynamo also shows reversals. It is shown here that laboratory and paleomagnetic data are both characterized by the presence of long-range correlations. 
\end{abstract}

\maketitle

\section{Introduction}
\label{Intro}

The observation of the paleomagnetic data \cite{generale,core,CK95} have shown that, unlike the solar magnetic field, where the polarity reversals are strictly periodic, geomagnetic measurements of the last $160$ million years present rather sudden and occasional polarity reversals. 
The reversal process is normally very rapid with respect to the typical time interval between successive reversals, which may range from $10^4$ up to $10^7$ years \cite{generale,CK95,valet93}.
Recent works on data analysis, experimental dynamo and theoretical modeling have inproved the knowledge of the Earth dynamo. However, the main fundamental questions concerning the polarity reversals still remain unanswered \cite{generale,review,dynamo,stefani05}. The nature of the triggers (external or internal to Earth) and the physical mechanisms giving rise to the reversals, the reason for the long time variations in the average reversal rate (cf. e.g. \cite{core,yamazaki}), are still open problems.

\begin{figure}
\centerline{\includegraphics[width=10.0 cm]{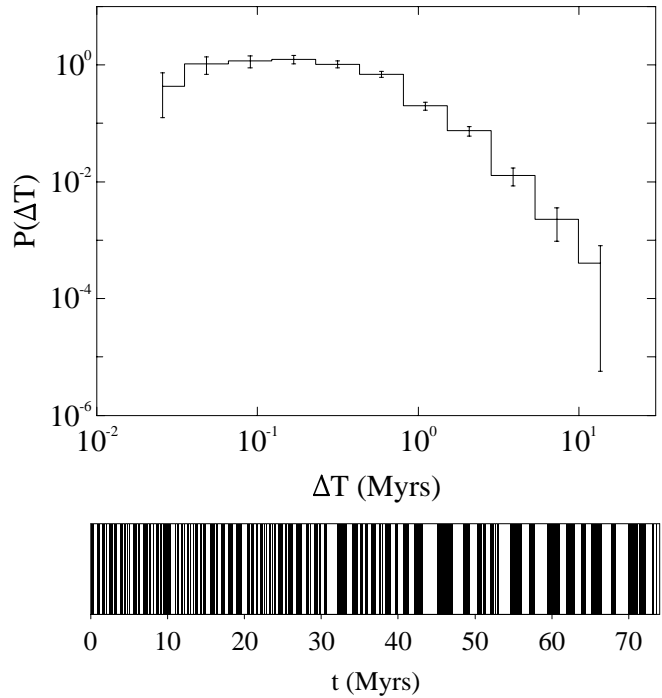}}
\caption{Bottom: Polarity of the earth's magnetic field (from today) as in the CK95 record (partial). 
The black bars are the normal (present) polarity. Top: the probability density function $P(\Delta t)$ of persistence times $\Delta t$ for CK95 database (statistical errors are shown as vertical bars).} 
\label{fig1}
\end{figure}

The sequence of geomagnetic reversals (see the example from the CK95 database \cite{CK95} shown in Fig. \ref{fig1}) seems to result from a of a stochastic process. The same behaviour is observed for experimental dynamo \cite{bourg} and from numerical simulations \cite{stefani05}. While experimental dynamo is a recent excellent achievement, the numerical approach, namely the direct solution of the Maghetohydrodynamics (MHD) equations (see \cite{review,earth1,earth2}) is still far from being satisfactory for a statistical analysis.
However, reversals are also observed in field resulting from simplified models, such as few modes models \cite{rikitake,crossley,turcotte}, models of noise-induced switchings between two metastable states \cite{schmitt,hoyng02,hoyng04}, or mean-field dynamo models with a noise-perturbed $\alpha$ profile \cite{stefani05}.

Recently, it has been shown through a simple statistical analysis, that the reversals of the paleomagnetic field are not random \cite{prl,pepi,gafd}, namely the statistics of interevent times ($\Delta t = t_{i+1} - t_i$, where $t_i$ is the time of the $i$-th event of the record) departs from a Poissonian distribution (namely an exponential law $P(\Delta t) = \lambda \exp(- \lambda \Delta(t))$, where $\lambda$ represents the reversal occurrence rate \cite{generale,hoyng02,mcfadden}), including a non-stationary Poisson process, in which case a power-law distribution could arise from the superposition of Poisson distributions with time variable rates $\lambda(t)$, see \cite{constable}. This result shows that geomagnetic reversals are clustered in time, probably because of presence of memory in the process generating polarity reversals. 

Here we show that experimental dynamo reversals also are characterized by correlations and clustering, suggesting that the reversal process is a universal property of dynamo, which does not need any external triggering.

\section{Local Poisson hypothesis and paleomagnetic data}
\label{LPH}

In this section we will describe the statistical tool used in this work to test, as a zero-th order hypothesis $H_0$, whether the observed sequence is consistent with a \textit{Local Poisson Process}. 
The reversals rate profile $\lambda(t)$ being in principle unknown, the test should be independent on it. A method introduced in cosmology \cite{bi} and more recently used for solar flares \cite{boffetta,lepreti} geomagnetic activity \cite{lepreti04}, random lasers in liquid crystals \cite{sameh}, and stock market analysis \cite{greco} will be used here. Consider the time sequence of reversals as a point-like process, and suppose that each reversal occurs at a discrete time $t_i$. The suitably normalized local time interval $h$ between reversals can be defined by introducing $\delta t_i$ as 
\begin{equation}
\delta t_i = \min \{t_{i+1}-t_i;t_i-t_{i-1}\} \; ,
\end{equation}
and $\tau_i$ by
\begin{equation}
\tau_i = \left\{\begin{array}{l}
t_{i-1}-t_{i-2} \;\;\;\;\;\;\;\;\;\;\; \mbox{if  } \delta t_i = t_i-t_{i-1} \\ 
t_{i+2}-t_{i+1} \;\;\;\;\;\;\;\;\;\;\; \mbox{if  } \delta t_i = t_{i+1}-t_i
\end{array} \right.
\label{taui}
\end{equation}

$\delta t_i$ and $\tau_i$ are then the two persistence times following or preceeding a given reversal at $t_i$. If the local Poisson hypothesis $H_0$ holds, both $\delta t_i$ and $\tau_i$ are independently distributed according to an exponential probability density: $p(\delta t) = 2 \lambda_i \exp(-2 \lambda_i \delta t)$ and $p(\tau)= \lambda_i \exp(-\lambda_i \tau)$ with local rate $\lambda_i$. 
The distribution of the variable $h$ defined by
\begin{equation}
h(\delta t_i, \tau_i) = \frac{2 \delta t_i}{2 \delta t_i + \tau_i}
\label{acca}
\end{equation}
will not depend on $\lambda_i$. 

For the surviving function of the probability density
\begin{equation}
P(h \geq H) = \int_H^{\infty} P(h) dh = \int_0^{\infty} dx 2\lambda e^{-2\lambda x} \int_0^{g(x,H)} dy \lambda e^{-\lambda y} 
\label{cumulativa}
\end{equation}
where $P(h)$ is the probability density function of $h$ and 

\[
g(x,H) = 2x \left[\frac{1}{H}-1\right] \; ,
\]

it can be easily shown that, under the hypothesis $H_0$,
\[
P(h \geq H) = 1-H \; ,
\]
that is, $h$ is a stochastic variable uniformly distributed in $h \in [0;1]$.

In a process where $\tau_i$'s are systematically smaller than $2 \delta t_i$'s, clusters are present and the average value of $h$ is greater than $1/2$. On the contrary, when the process is characterized by voids, the average value of $h$ is less than $1/2$. From time series, it is easy to calculate the surviving function $P(h \geq H)$ and the probability density function $P(h)$. 

The test described above has been recently applied to four different datasets of geomagnetic polarity reversals, including the already mentioned CK95 cite{prl,pepi,gafd}. The probability density function $P(h)$ is reported in Fig. ref{fig2} for the CK95 datasets. A significant deviation from the uniform distribution was observed in all the datasets, due the presence of clusters.

\begin{figure}
\centerline{\includegraphics[width=10.0 cm]{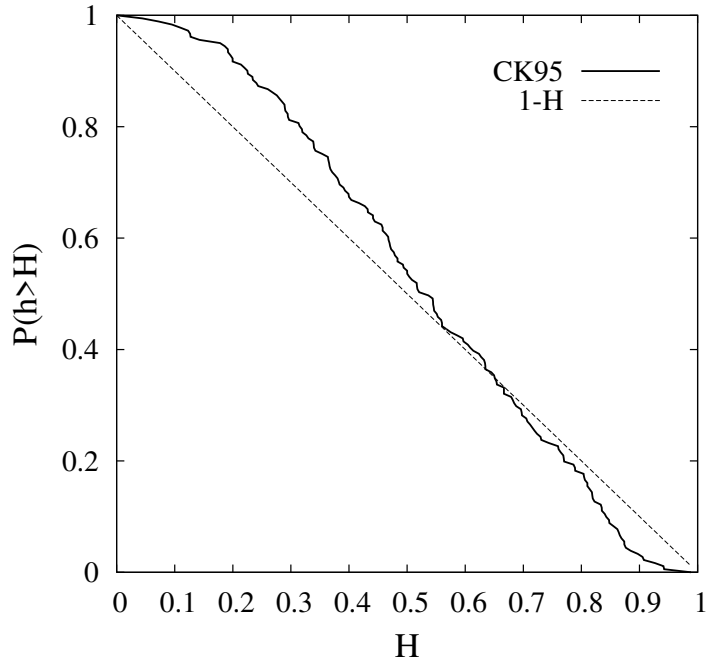}}
\caption{Probability densities $P(h)$ of the stochastic variable $h$ and corresponding surviving functions $P(h\geq H)$  for all the empirical datasets. The theoretical probability expected under a Poisson statistics is also shown.}
\label{fig2}
\end{figure}

\section{Experimental dynamo}

\begin{figure}
\centerline{\includegraphics[width=8.5cm]{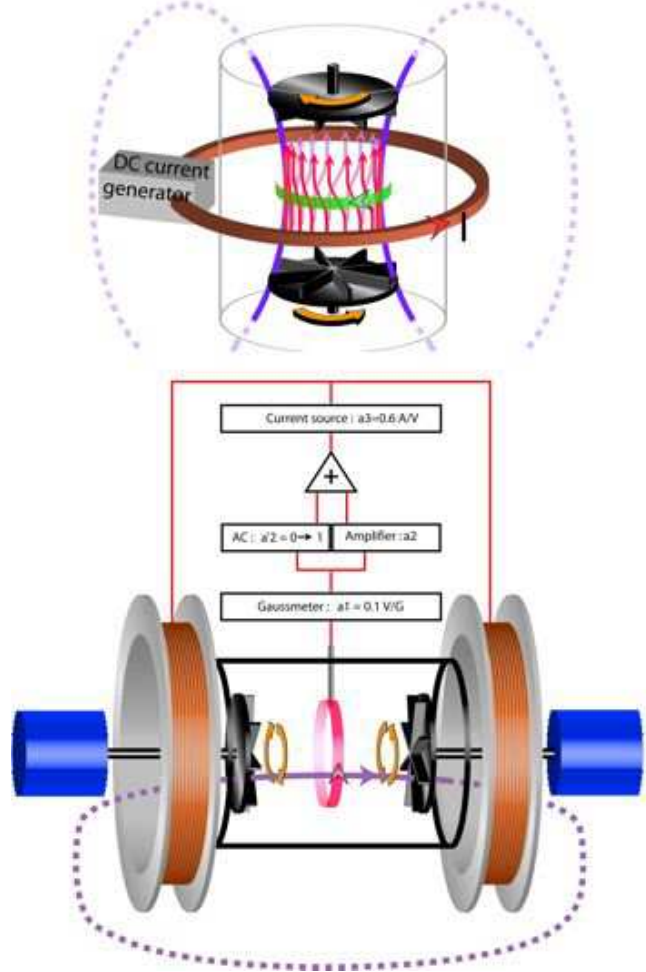}}
\caption{(a) Omega effect : the differential rotation on a von K\'arm\'an flow advects and stretches an externally applied axial field $B_{0z}$ so as to generate a toroidal component $B_\theta$. (b) Postive feed-back : the amplitude of $B_\theta$ is used to drive a power source which generates the current in the external loop. Two Helmoltz coils are set on either end of the cylindrical flow vessel;  $B_\theta$ is measured in the mid-plane by a Hall probe connected to a Bell gaussmeter. The measured value is fed into a linear amplifier whose output drives a Kepco current source. In order to explore the role of the turbulent fluctuations, the amplifier has separate channels for the DC and flcutuating parts of the induction.
}
\label{setup} 
\end{figure}
The dynamo laboratory model \cite{bourg} mimics an alpha-omega cycle where part of the dynamo cycle is generated by an external feed-back but the flow turbulence is still included and has a leading role.  In order to achieve this in a simple laboratory dynamo, we relax the requirement that the current path be fully homogeneous, and we effectively prescribe an alpha mechanism by which a toroidal magnetic field generates an induced poloidal one. However, the omega poloidal to toroidal conversion still results from a fully turbulent process. Our experimental fluid turbulent dynamo is very much inspired by a variation of the solid rotor dynamo proposed by Sir Edward Bullard in the early 20th century, and described in figure 1. Two coaxial disks counter rotate at a rate $\Omega$. When an axial magnetic field $\mathbf{B}_{0z}$ is externally applied, the flow differential rotation induces a toroidal field $\mathbf{B}_\theta$, this is the omega effect. The value of this field is then used to drive a linear current amplifier in the loop that generates $\mathbf{B}_{0z}$. The poloidal to toroidal conversion is entirely due to the fluid motion, and incorporates all turbulence effects. It has been extensively studied in previous ``open loop'' experiments. When $\mathbf{B}_{0z}$ is externally fixed, one has $B_\theta = k R_m B_{0z}$ where $R_m = R^2\Omega/\lambda$ is the magnetic Reynolds number (with $\lambda$ the magnetic diffusivity of liquid Gallium) and $k$ is a ``geometric'' constant which in our experiment has been measured of the order of 0.1. The toroidal to poloidal conversion is then obtained by feeding the axial coils with an electrical current linearly driven by a signal proportional $B_{1\theta}$, so that $B_{0z} = \alpha G B_\theta$ which reinforces $B_{0z}$ with $G$ an adjustable gain. In such a closed loop setup, one then has $B_z = \alpha G k R_m B_{0z}$, and self-sustained dynamo is reached as $\Omega > \Omega^c = \lambda / GkR^2$. Clearly, the adjustable gain of the linear amplifier allows to adjust the value of $\Omega^c$ to an experimentally accessible range. At this point it should be emphasized that although the feed-back scheme is very similar for the Bullard rotor dynamo and for our fluid experiment, the expected dynamics is much richer because of the strong fluctuations in the turbulent flow, where Reynolds numbers in excess of $10^6$ are reached. Indeed, the von K\'arm\'an flow is known for its complex dynamics, presenting not only small scale turbulent fluctuations but also large scale ones -- for instance fluctuation up to 114\% for the differenteial rotation effect has been reported). Compared to the 1963 pioneering experiment of Lowes and Wilkison with solid rotor motions, the study here fully incorporates fluid turbulence and the associated fluctuations of magnetic induction. The role of these fluctuations, inherent to large Reynolds number flows, remains one of the mysterties of natural dynamos, and of noisy instabilities in a braoder framework.

In this experiment the value of the magnetic field at saturation $B_{\rm sat}$  is fixed by the maximum current that can be drawn from the power amplifier driving the coils. We measure $B_{\rm sat} \sim 30$~G, a value such that the Lorentz forces cannot modify the hydrodynamic flow --since it yields an interaction parameter of the order of $10^{-3}$. The saturation of the instability is therefore driven by the amplifier non-linearities rather than by the back-reaction of Lorentz forces on the dynamical velocity field. As a consequence, the $B_z$ component of the generated magnetic field saturates at the same mean amplitude $B_{\rm sat}$ for all rotation rates ($B_{\rm sat}$ corresponds to the magnetic field generated by the coils when the current source is saturated), the saturation amplitude of the toroidal field $B_{\theta \rm sat}=k R_m B_{\rm sat}$ linearly increases with $\Omega$. 

Another noteworthy observation is that the presence of turbulent fluctuations plays a crucial role in the triggering of the magnetic field reversals. In the experimental results reported here, the current source is driven by an amplifier whose input is $\overline{B}_\theta + g.b'_\theta$, with $\overline{B}_\theta$ the low pass DC component of $B_\theta$ and $b'_\theta$ its AC fluctuating part. This arrangement allows to study separately the role of slow variations and turbulent fluctuations in the feed-back loop. In the results reported in this article, we have set  $g=1.18$. A homopolar dynamo, i.e. without reversals, was obtained for smaller values of $g$ or when the $b'_\theta$ input in the amplifier was replaced by a synthetic gaussian white noise (even with a high amplitude).

We show here the results of the $h$-test obtained in a realization (serie27) with $\Omega=12$ Hz, and cutoff frequency $f_c=600$ mHz.  Similar resutls were observed with different parameters, and this study is left for more extended work.
Figure ref{fig3} shows the reversals surviving function in the case described here. The behaviour is very similar to the paleomagnetic case, indicating again presence of clustering and correlations, rather than a random behaviour. This indicates that the mechanism responsible for the clustering is present in both dynamoes, suggesting some sort of universality of the process.

\begin{figure}
\centerline{\includegraphics[width=10cm]{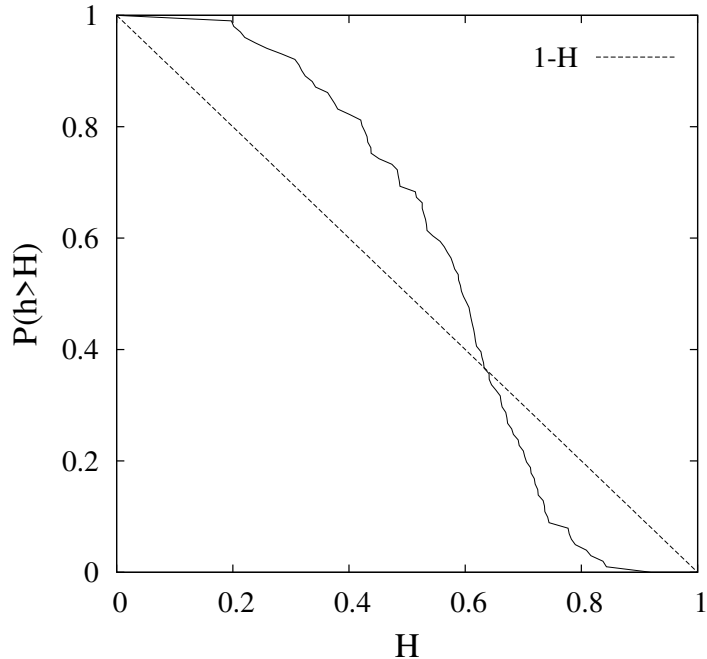}}

\caption{Probability densities $P(h)$ of the stochastic variable $h$ and corresponding surviving functions $P(h\geq H)$  for the experimental dataset described in the text. The theoretical probability expected under a Poisson statistics is also shown.}
\label{fig3}
\end{figure}

\section{Conclusion}
\label{fine}

In this short paper, the statistical properties of persistence times between geomagnetic reversals have
been investigated. We performed a statistical test which showed that geomagnetic reversals are produced
by an underlying process that is far from being locally Poissonian, as recently conjectured by \cite{constable}. 
Thus, the sequence of geomagnetic reversals is characterized by time correlations. As spontaneous reversals of the geodynamo field have been observed in high resolution numerical simulations \cite{earth1,earth2}, the main results contained in this paper seem to indicate that such reversals could be related to the non-linear nature of the turbulent dynamo. 
In order to confirm this conjecture, we performed the statistical test mentioned above on recent results from laboratory dynamo.
Our analysis has shown that the departure from Poisson statistics found in the paleomagnetic data, related with the long range
correlations introduced by the chaotic dynamic of the system cite{pepi,gafd}, are also present in the laboratory dynamo.
Such correlations can be associated with the presence of some degree of memory in the underlying dynamo process \cite{valet,stefani05} which gives rise to clustering of reversals.

\end{document}